\documentclass[twocolumn,superscriptaddress,amsmath,amssymb,prl,aps]{revtex4}
\usepackage{epsfig}
\usepackage{graphicx}
\usepackage{psfrag}
\usepackage{subfigure}
\begin{document}

\title{Untangling influences of hydrophobicity on protein sequences and structures}

\author{Mehdi Yahyanejad
%\email[Correspondence to:~{\em mehdi@mit.edu}]
}
\affiliation{Department of Physics, Massachusetts Institute of Technology, Cambridge,  MA 02139}
\affiliation{Department  of Biology, Massachusetts Institute of Technology, Cambridge, MA 02139}
\author{Christopher B. Burge}
\affiliation{Department  of Biology, Massachusetts Institute of Technology, Cambridge, MA 02139}
\author{Mehran Kardar}
\affiliation{Department of Physics, Massachusetts Institute of Technology, Cambridge,  MA 02139}

%\date{\today}

\begin{abstract}
We fit the Fourier transforms of solvent accessibility and
hydrophobicity profiles of a representative set of proteins to a
joint multi-variable Gaussian. This allows us to separate the
intrinsic tendencies of sequence and structure profiles from the
interactions that correlate them;  
for example, the
$\alpha$-helix periodicity in sequence hydrophobicity is dictated by
the solvent accessibility of structures.  
The distinct intrinsic  tendencies of sequence and structure profiles 
are most pronounced at long periods, where sequence hydrophobicity 
fluctuates more, while solvent accessibility fluctuations are less than average.
Interestingly, correlations between the two profiles can
be interpreted as the Boltzmann weight of the solvation energy at room temperature.
\end{abstract}

\maketitle
\clearpage

%\section*{Introduction}
How the sequence of amino acids determines the structure and function
of the folded protein remains a challenging problem.  
It is known that hydrophobicity is an important determinant of the 
folded state; hydrophobic monomers tend to be in the core, and polar
monomers on the surface~\cite{kauzmann59, lee71,eisenberg84, moelbert04}.  
Several studies have examined the correlations in the hydrophobicity of 
amino-acids along the protein chain~\cite{irback96, pande94, strait95, weiss98},
which are in secondary structure prediction~\cite{eisenhaber96a}, 
and in the design of good folding sequences~\cite{wilder00}. 
Naturally, sequence correlations arise from locations of the
amino-acids in the folded protein structure, and %hydrophobicity profiles
are best interpreted in conjunction with solvent accessibility profiles (which
indicate how exposed a particular amino-acid is to water in a specific structure).
%Previous work, however, has not explicitly accounted for  the impact of solvent accessibility.
For example, Eisenberg {\em et  al}~\cite{eisenberg84} note
that for secondary structures lying on the protein surface, 
which have a strong periodicity in their solvent accessibility, 
hydrophobicity profiles also exhibit the period
of the corresponding $\alpha$ helix or $\beta$ strand. 
Constraints from forming compact structures induce 
strong correlations in the solvent accessibility
profile~\cite{govorun01,strait95,yahyanejad03,khokhlov99}, which should in
turn induce similar correlations in the hydrophobicity profiles.
It is desirable to quantify and separate the resulting correlations in protein
sequences and structures.

In this paper, we aim for a unified treatment of hydrophobicity and
solvent accessibility profiles, and the interactions between them. 
The {\em sequence} of each protein is represented by a
profile $\left\{h_i\right\}$, where $h_i$ is a standard measure of the
hydrophobicity of the $i$-th amino-acid along the backbone~\cite{biswas03}.  
Its {\em structure} has a profile $\left\{s_i\right\}$ for $i=1,~2,~\cdots,N$, 
where $s_i$ is a measure of the exposure of the amino-acid to water 
in the folded structure~\cite{lee71}.  
While we do not expect perfect correlations
between these profiles, we can inquire about the statistical nature of
these correlations, and in particular whether they are diminished or
enhanced at different periods.  To this end, we employ the method of
Fourier transforms , and examine the
statistics of the resulting amplitudes $\{\tilde{h}_q,\tilde{s}_q\}$,
and power spectra $\{|\tilde{h}_q|^2,|\tilde{s}_q|^2\}$, for a
database of 1461 non-homologous proteins.
In a sense, this can be regarded as extending the
work of Eisenberg {\em et al}~\cite{eisenberg84} who
explore correlations between hydrophobicity and solvent
independent of specific locations along the backbone.
Of course, the use of Fourier analysis is by no means new, and
has for example been employed to study hydrophobicity 
profiles~\cite{eisenberg84, rackovsky98, irback96,irback00}.  
However, we are not aware of its use as a means of correlating sequence and
structure profiles.
  
Our results suggest that the hydrophobicity and solvent accessibility profiles
are well approximated by a joint Gaussian probability distribution.  
This  allows us to obtain the {\it intrinsic} correlations in the hydrophobicity profile, 
as distinct from correlations induced by solvent accessibility.  
For example,  the $\alpha$-helix periodicity in hydrophobicity profiles is shown to be
induced by the corresponding periodicity in the solvent accessibility profiles. 
We  also find that at long wavelengths the two 
profiles have different intrinsic characteristics: 
solvent accessibility profiles are positively correlated
while hydrophobicity profiles are anti-correlated.  
Interestingly, the coupling between the two profiles is independent of wave-number,
and hence can be interpreted as the Boltzmann weight of the solvation energy. 
The corresponding temperature is close to room temperature,
consistent with the ``mean'' temperature estimated in previous work
from the frequencies of occurrence of amino acid residues in the core
and on the surface~\cite{miller87, finkelstein95}.

For our protein data set, we selected 2200 representative chains
from the Dali/FSSP database. Any two protein chains in this set
have more than 25 percent structural dissimilarity. We removed all the 
multi-domain chains by using the CATH domain definition database, leaving
 1461 protein chains~\cite{holm98b, orengo97, moelbert04}. 
The hydrophobicity profiles, $\{h_i\}$, were generated from the sequence of amino-acids
using the experimentally measured scale of  Fachere and Pliska~\cite{fauchere83}
(in units of kcal/mol).
We used the {\em relative solvent accessibility} 
reported by NACCESS~\cite{hubbard91} to generate solvent accessibility profiles $\{s_i\}$.
(The relative solvent accessibility is the ratio of the solvent accessibility of a residue
 to the solvent  accessibility of that residue in an extended tripeptide
 ALA-X-ALA for each amino acid type X.)
%Our result was in consistant when we used a consensus hydropathy
%scale calculated by Carugo ~\cite{carugo03}.  Solvent accessibilities
%$\{s_j\}$ do not have any units because we used the relative solvent accessibility. 
We then computed the corresponding Fourier components as
%calculated the Fourier transform of the solvent accessibility and hydrophobicity profiles to generate the power spectra. For each structure vector $\{s_j\}$, the Fourier components were obtained from
\begin{equation}
\label{sq}
\begin{array}{c}
\tilde{s}_q = {\frac{1}{\sqrt{N}}}\sum_{j=1}^N e^{iqj} \left(s_j-\frac{\sum_{j=1} s_j}{N}\right) ,
%  \text{\rm power spectrum }= \vert s_q \vert^2 \\
\end{array}
\end{equation}
where $q=2\pi\alpha/N$, with $\alpha = 0,1,\cdots,N-1$, and similarly for $\tilde{h}_q$.
(The average values were subtracted to remove the DC  component in the Fourier transform.)

%The corresponding power spectrum is $|\tilde{s}_q|^2$.  Since proteins
%are in different lengths, each $|\tilde{s}_q|^2$ has different number
%points in the interval of $[0,\pi]$.  Therefore to find the average
%power spectrum, we put all calculated points in $50$ bins in the
%interval of $[0,\pi]$, and then took the average of ${\vert
%  \tilde{s_q} \vert^2}$ in each bin by noting that a point belonging
%to a protein of length $N$ should be given a weight of
%$\frac{100}{N}$, which equalizes the contribution of short and long
%proteins to average power spectrum.  As a check, we estimated the
%average value for our power spectra through Parseval's theorem $\sum_q
%|\tilde{s}_q|^2 = \sum_i (s_i-\overline{s_i})^2 $. Here
%$\overline{s_i}$ is the average solvent accessibility in each protein
%structure. From histograms of $s_i - \overline{s_i} $, we found
%$<(s_i-\overline{s_i})^2> = 0.1$; from Fig.~\ref{Sq2}, we saw that the
%average $|\tilde{s}_q|^2$ was indeed the same. This analysis was
%repeated for hydrophobicities, giving $<(h_i-\overline{h_i})^2> = 0.96
%\rm{kcal^2/mol^2} $, which is indeed the average of $|\tilde{h}_q|^2$
%as in Fig.~\ref{Hq2}.
\begin{figure}[htbp]
\centering 
\renewcommand{\subfigcapskip}{2pt}
\renewcommand{\subfigbottomskip}{2pt}
\subfigure[]{\label{Sq2}
\psfrag{X}[l][][1.3][0] {$q$}      
\psfrag{Y}[bc][][1.3][0]{$<|\tilde{s}_q|^2>$}
\psfrag{L}[l][][1.3][0] {$\lambda$}  
\epsfig{file=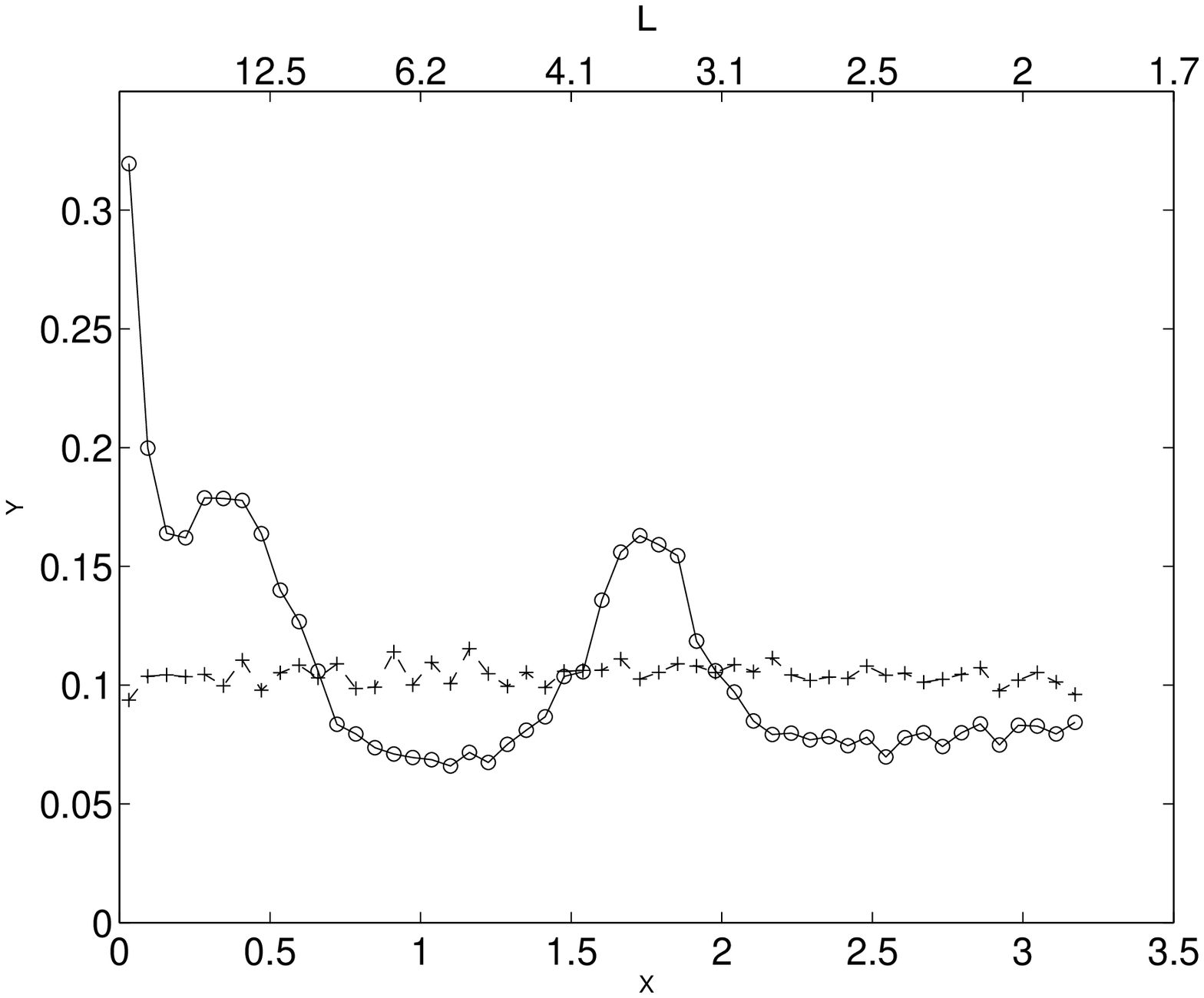, height=1.75in}}
\subfigure[]{\label{Hq2}
\psfrag{X}[l][][1.3][0] {$q$}      
\psfrag{Y}[bc][][1.3][0]{$<|\tilde{h}_q|^2>$}
\psfrag{L}[l][][1.3][0] {$\lambda$}  
\epsfig{file=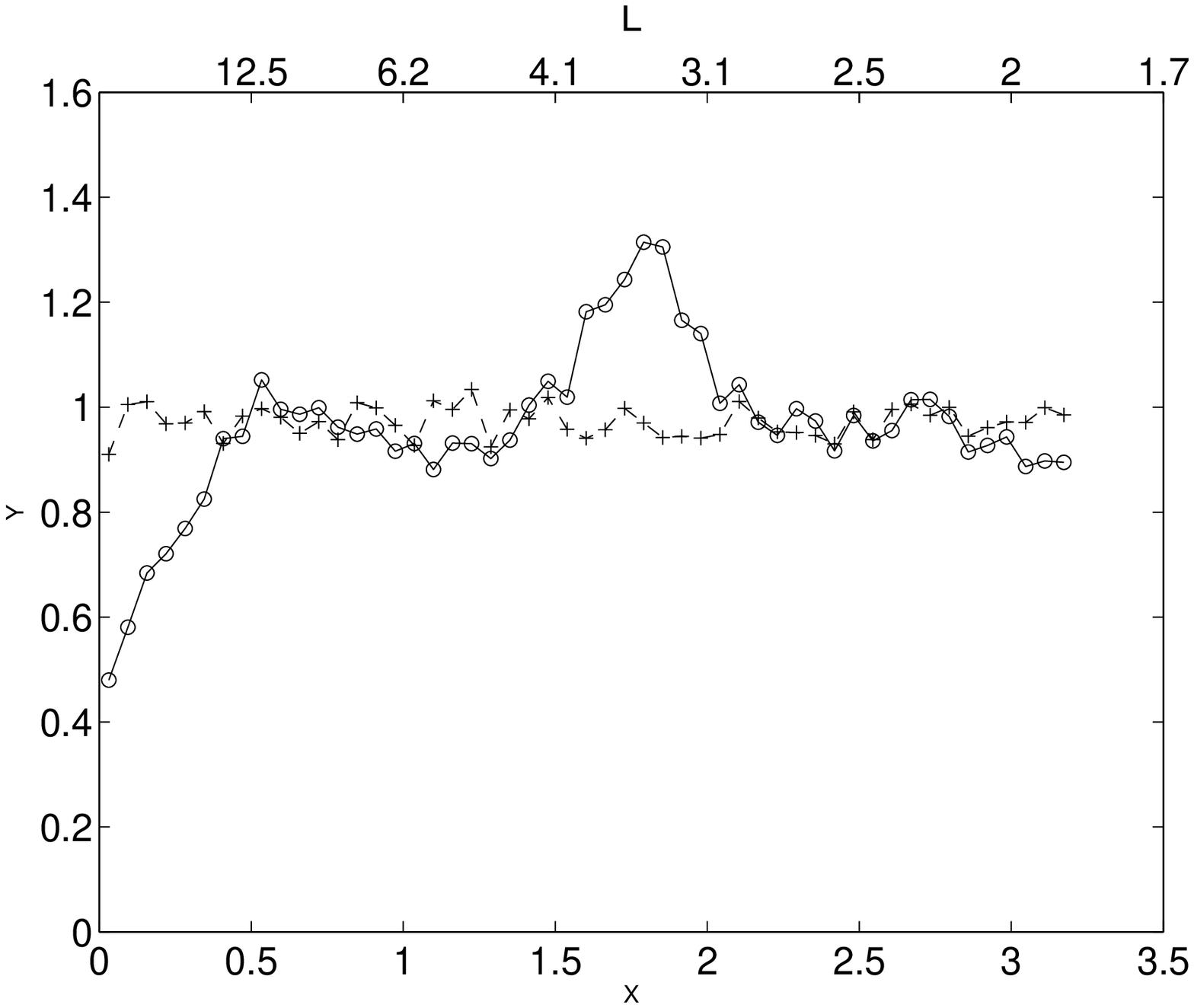, height=1.75in}}
\caption{Power spectra from averaging over 1461 proteins, for
(a) solvent accessibility (there are no units for $|\tilde{s}_q|^2$,
since it based on accessibility relative to solution);
(b) hydrophobicity (the units of  $|\tilde{h}_q|^2$ are (k~cal/mole)$^2$).
The plus signs in each case are obtained from random permutation of the sequences.
}
\label{powerspectra}
\end{figure}

Our results for the power spectra of solvent accessibility and
hydrophobicity profiles are indicated respectively in Figs.~\ref{Sq2}
and \ref{Hq2}
($q$ is related to the periodicity $\lambda$ through $\lambda =\frac{2\pi}{q}$).  
A prominent feature of both plots is the peak at the
$\alpha$-helix periodicity $\lambda=3.6$~\cite{eisenberg84}.
Its presence in the solvent accessibility spectrum indicates that
solvation energy plays a role in the spatial arrangement of $\alpha$
helices when they lie on the surface, the hydrophobic
monomers are more likely to be exposed to the solvent. ~\cite{eisenberg84}.

\begin{figure}[htbp]
\centering 
\renewcommand{\subfigcapskip}{2pt}
\renewcommand{\subfigbottomskip}{2pt}
\subfigure[]{\label{covariance_q1_q2}
\psfrag{X}[l][][1.3][0] {$q$}      
\psfrag{Y}[bc][][1.3][0]{$q^\prime$}
\psfrag{T}[l][][1.3][0] {$<\Re s_q \Re s_{q^\prime}>$}  
\epsfig{file=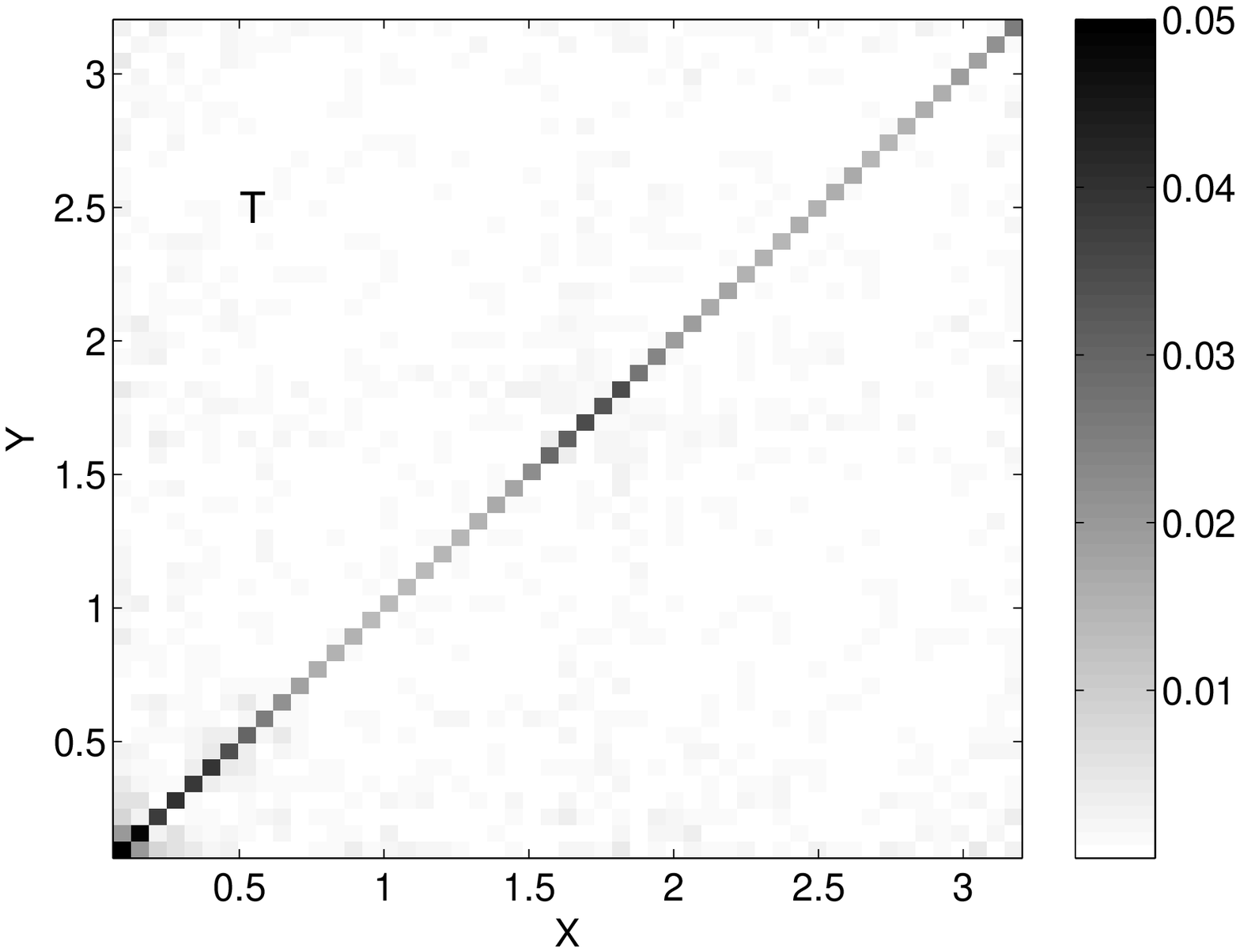, height=1.75in}}
\subfigure[]{\label{Sq_vs_Hq}
\psfrag{X}[tl][][1.3][0] {$\Re s_q$}      
\psfrag{Y}[c][][1.3][0]{$\Re h_q$}
\epsfig{file=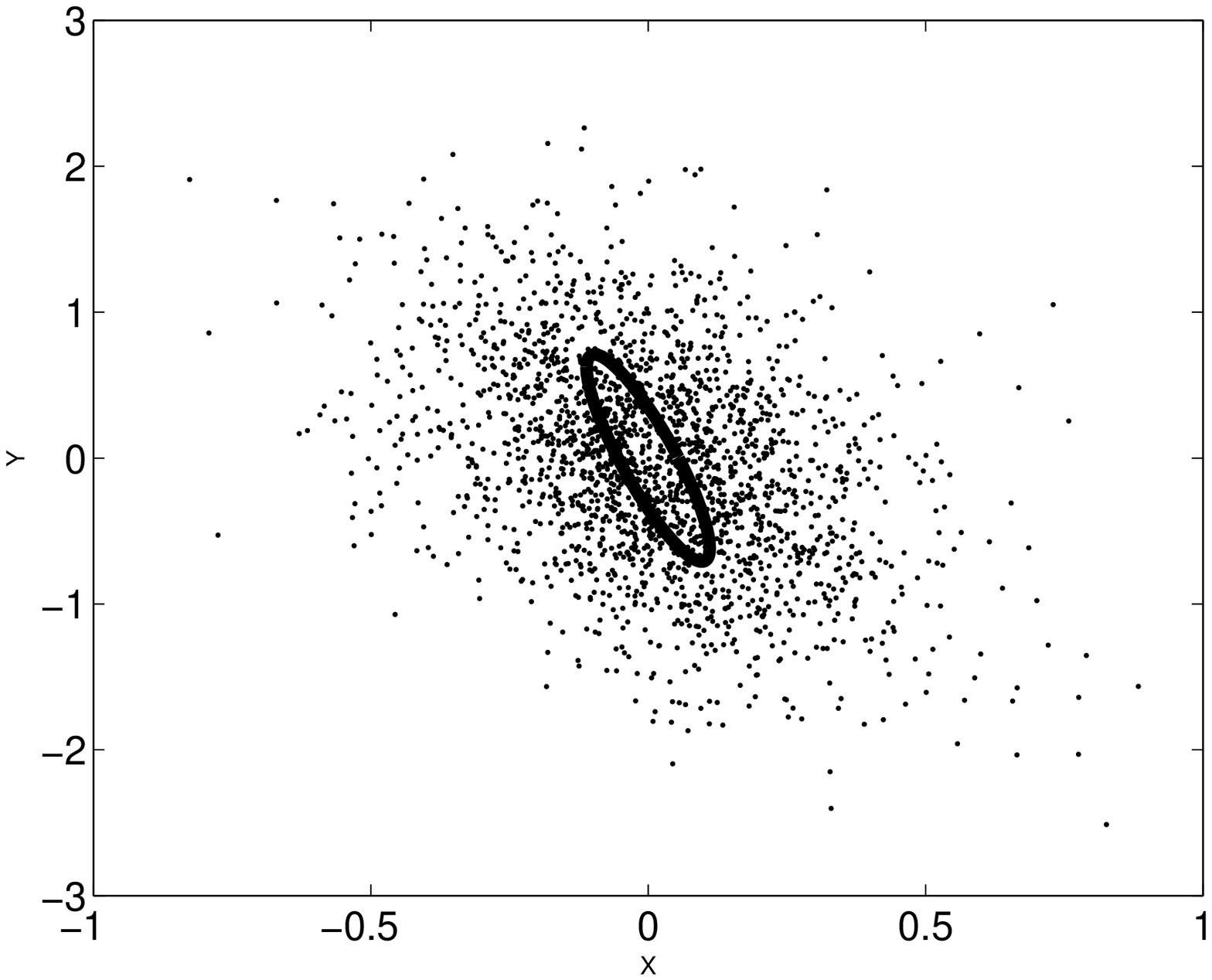, height=2in}}
\caption{
(a) Covariance of $\Re s_q$ with  $\Re s_{q^\prime}$. There is very little
correlation between off-diagonal terms.
(b) Scatter plot of $\Re h_q$ versus $\Re s_q$ for $q=0.90$, and the half-width half-maximum
locus of a Gaussian fit  (solid line).
}
\label{correlations}
\end{figure}

We would like to untangle correlations between the two profiles, so as
to determine their intrinsic tendencies, by finding a joint
probability distribution $P(\{s_i\},\{h_i\})$.  Clearly this cannot be
decomposed as a product of contributions from different sites $i$, as
neighboring components such as $s_i$ and $s_{i+1}$, are highly
correlated.  We anticipate that the Fourier components for different
$q$ are independently distributed
(i.e. $P(\{\tilde{h}_q,\tilde{s}_q\})=\prod_q
p(\tilde{h}_q,\tilde{s}_q)$) for the following reasons: 
{\bf (i)} For the subgroup of cyclic proteins~\cite{weikl03} the index $i$ is arbitrary, 
and the counting can start from any site. 
The invariance under relabeling then implies that the probability
can only depend on $i-j$, and hence separable into independent Fourier components.
This exact result does not hold for open proteins because of end effects, but should
be approximately valid for long sequences when such effects  are small.
{\bf (ii)} Numerical analysis of a lattice model of proteins in Ref.~\cite{yahyanejad03} 
confirms the exact decomposition into Fourier modes for cyclic structures,
and its robustness even for open structures of only $N=36$ monomers.
To test this hypothesis, we examine all possible covariances involving 
$\{\tilde{h}_q,\tilde{s}_q\}$ for different $q$. 
Note that the Fourier amplitudes are complex 
(i.e. $\tilde{s}_q=\Re s_q+i\Im s_q$, and similarly for $\tilde{s}_q$),
and hence there are $4\times 4$ covariance plots, such as in
Fig.~\ref{covariance_q1_q2}) for the covariance of $\Re s_q$ with itself.
In all cases we find that the diagonal terms are small; the only exceptions
are at small $q$ where we expect end effects to be most pronounced.

\begin{figure*}[!htbp]
\centering 
\renewcommand{\subfigcapskip}{2pt}
\renewcommand{\subfigbottomskip}{2pt}
\mbox{
\subfigure[]{\label{A_q}
\psfrag{X}[l][][1.3][0] {$q$}      
\psfrag{Y}[bc][][1.3][0]{$A_q$}
\psfrag{L}[l][][1.3][0] {$\lambda$}  
\epsfig{file=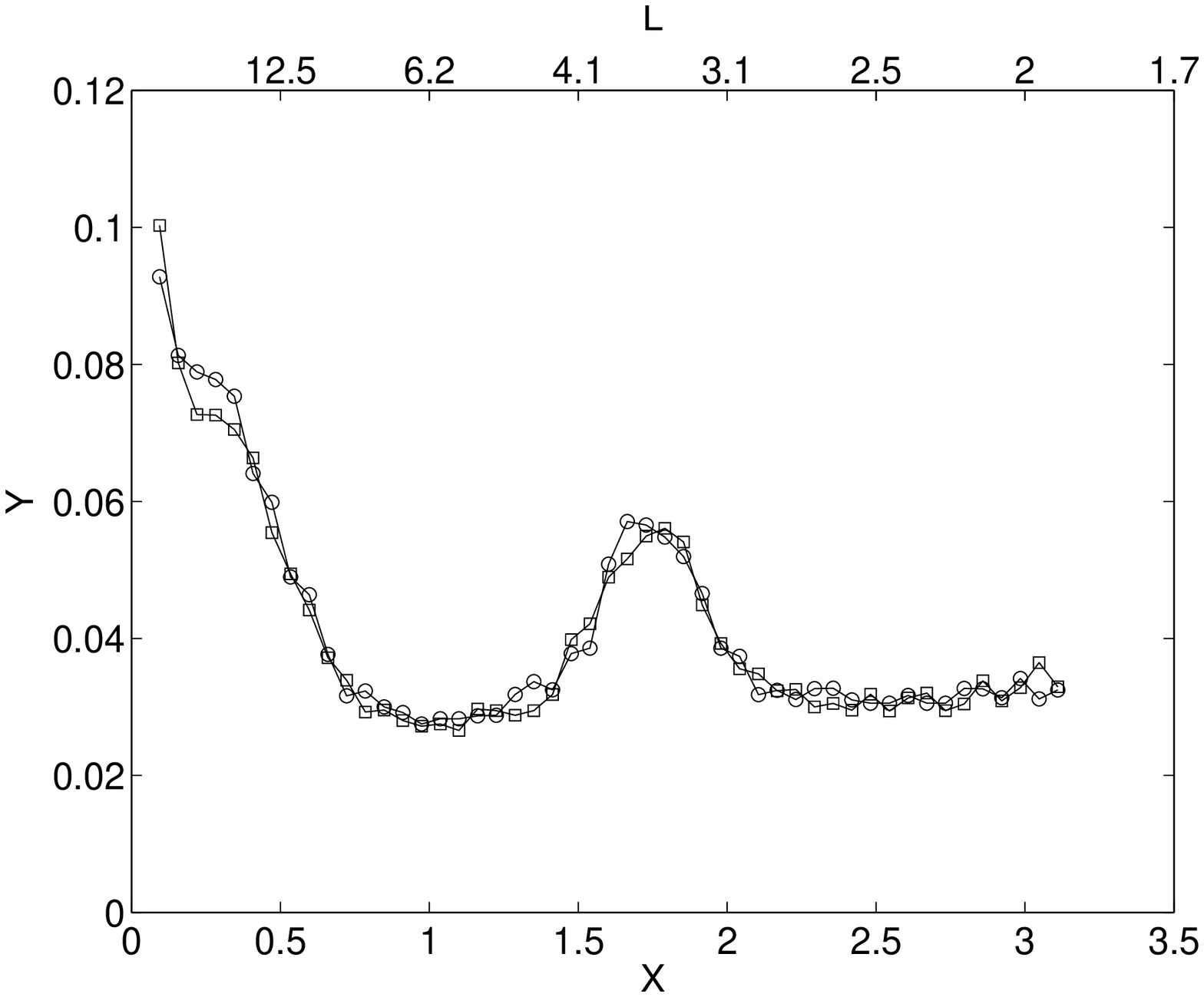, height=1.75in}
}\quad
\subfigure[]{\label{B_q}
\psfrag{X}[l][][1.3][0] {$q$}      
\psfrag{Y}[bc][][1.3][0]{$B_q$}
\psfrag{L}[l][][1.3][0] {$\lambda$}  
\epsfig{file=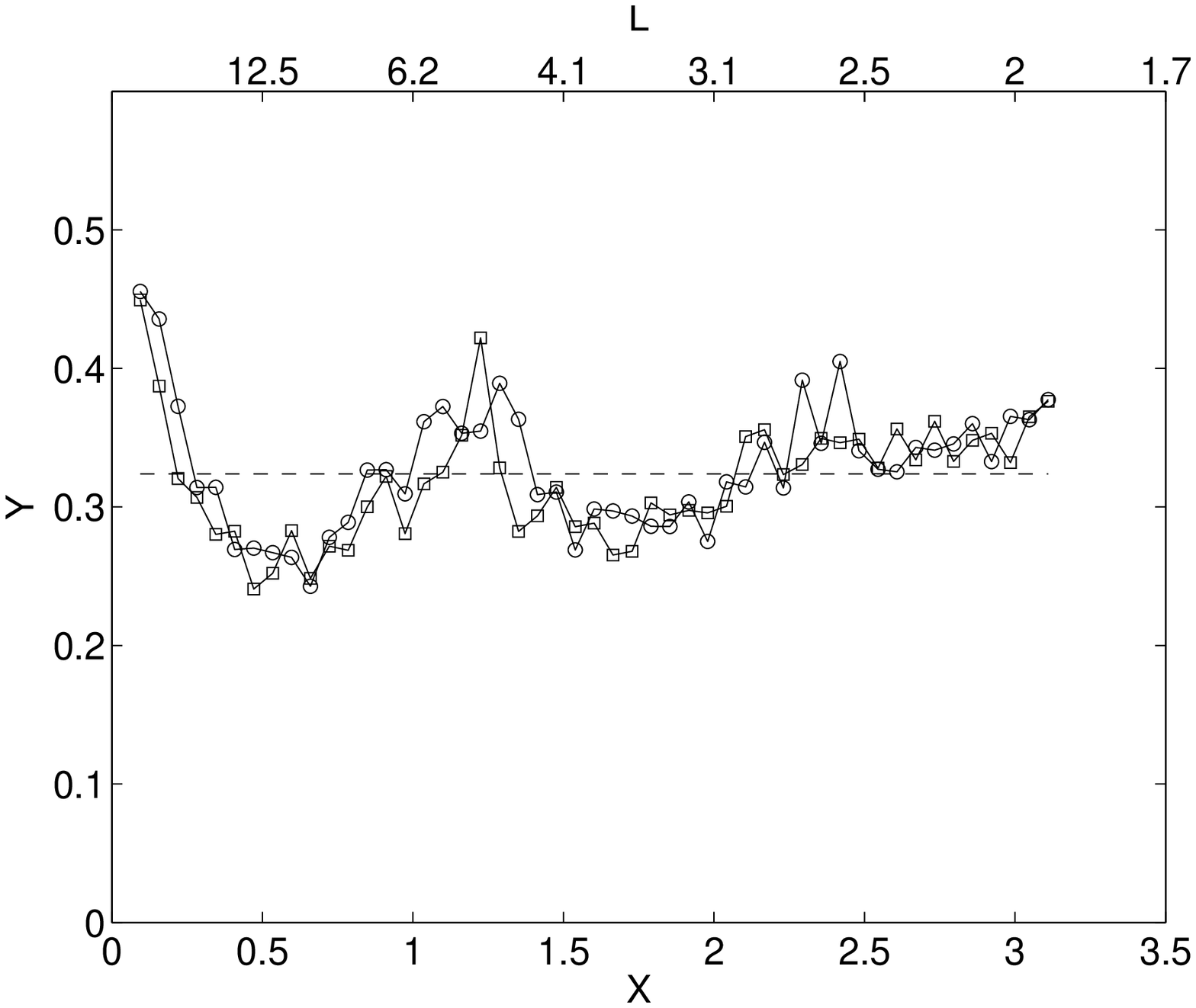, height=1.75in}
}\quad
\subfigure[]{\label{C_q}
\psfrag{X}[l][][1.3][0] {$q$}      
\psfrag{Y}[bc][][1.3][0]{$C_q$}
\psfrag{L}[l][][1.3][0] {$\lambda$}  
\epsfig{file=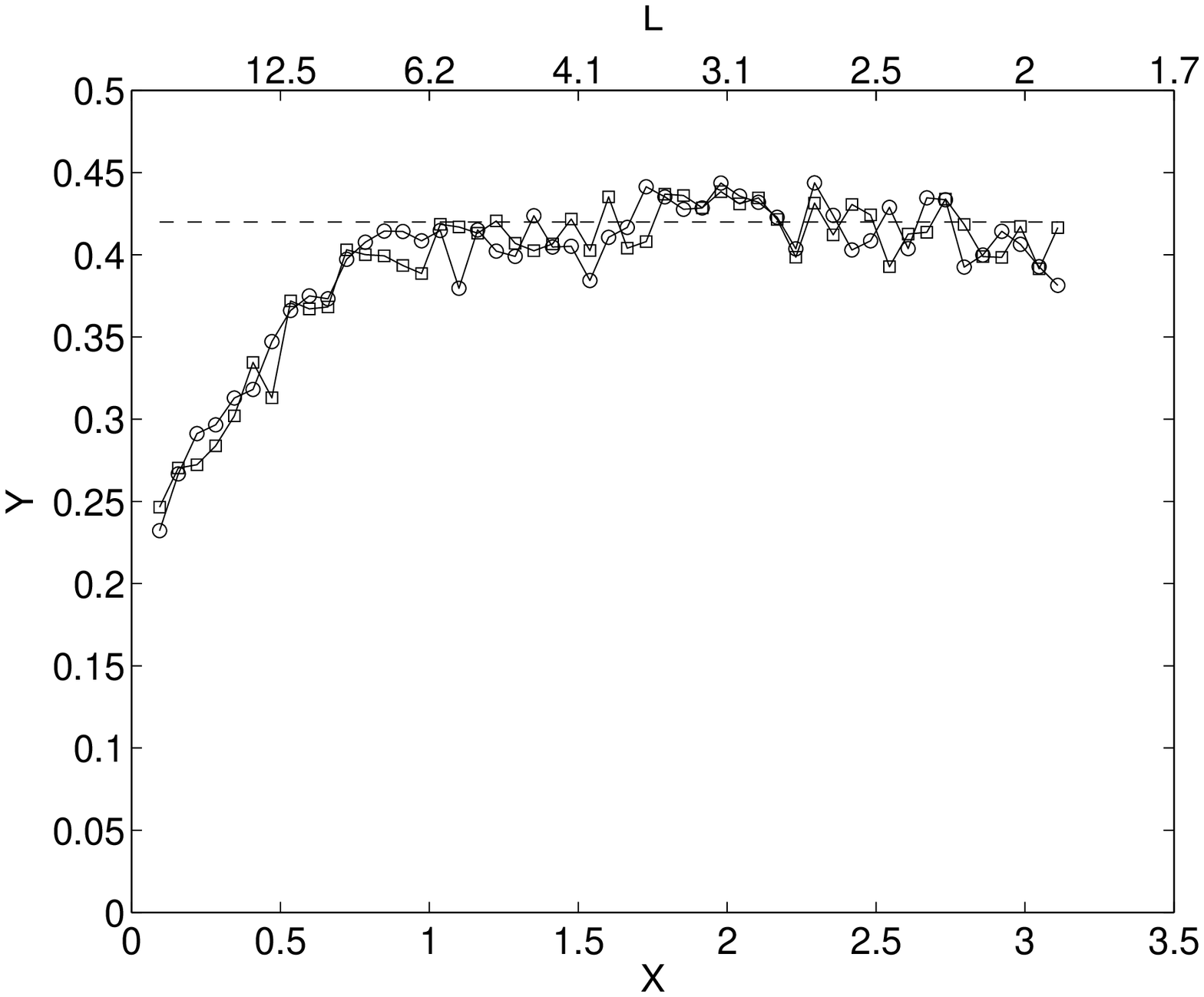, height=1.75in}
}
}
\caption{{\em Intrinsic} variances of solvent accessibility and hydrophobicity 
profiles are described by $A_q$ and $C_q$ respectively, while $B_q$
is related to the interaction that correlates them.
The square and circle symbols correspond to the parameters of the
imaginary and real components, respectively. These figures are calculated for our set of 1461 proteins.
Dashed lines indicate respectively  the average value of $B_q$ [in (b)], and 
the asymptotic behavior of $C_q$ [in (c)].} \label{pre_factors}
\end{figure*}

One can make a similar case for the independence of the real and imaginary
components at a given $q$. (For cyclic structures the phase is arbitrary.)
The real (imaginary) components are, however, correlated as illustrated
by the scatter plot of $(\Re h_q,\Re s_q)$ for $q=0.9$ in Fig.~\ref{Sq_vs_Hq}.
We made similar scatter plots for different values of $q$ in the interval 
$0$ to $\pi$, with similar results which were well fitted to Gaussian forms.
Based on these results, we describe the joint probability distribution in
Fourier space by the multivariate Gaussian form
\begin{eqnarray}
\label{eqn:Gaussian}
%P\left(\left\{\tilde{h}_q,\tilde{s}_q\right\}\right) 
P(\{\tilde{h}_q,\tilde{s}_q\}) 
&=& \prod_{q}
\exp\left[-\frac{(\Re s_q)^2}{2A_q} - \frac{\Re s_q\Re h_q}{B_q}- \frac{(\Re h_q)^2}{2C_q}\right] 
\notag \\ 
  &\times& 
 \exp\left[-\frac{(\Im s_q)^2}{2A'_q} - \frac{\Im s_q\Im h_q}{B'_q}- \frac{(\Im h_q)^2}{2C'_q}\right] ,
\end{eqnarray}
with the parameters plotted in Fig.~\ref{pre_factors}.
If the probabilities depend only on the separation $i-j$ between sites, the real and
imaginary Fourier amplitudes should follow the same distribution.
In our fits we allowed the corresponding parameters to be different
to obtain a measure of the accuracy of the model and the fitting procedure.
As indicated in Fig.~\ref{pre_factors} the resulting values are quite close,
differing by less than 5\%.

We interpret $\{A_q\}$ and $\{C_q\}$ as measures of {\em intrinsic}
tendencies of hydrophobicity and surface exposure profiles, while
$\{B_q\}$ indicates the strength of the interactions that correlate them.
In the absence of any such interactions, $\{A_q\}$ and $\{C_q\}$ 
would be the same as the power spectra in Fig.~\ref{powerspectra}.
With this in mind, let us now examine these plots in more detail.
%They are related to the original variables $<|s_q|^2>$ and $<|h_q|^2>$ through:%\begin{eqnarray}
%\label{eqn:Transform}
%<|s_q|^2> &=& \frac{A_q}{1 - \frac{A_q C_q}{B_q^2}}+ \frac{A'_q}{1 - \frac{A'_q C'_q}{{B'}_q^2}} \notag \\
%<|h_q|^2> &=& \frac{C_q}{1 - \frac{A_q C_q}{B_q^2}}+ \frac{C'_q}{1 - \frac{A'_q C'_q}{{B'}_q^2}}
%\end{eqnarray}
%In the absence of interactions, or $1/B_q\to0$, these equations reduce to 
%$<|\tilde{s}_q|^2>=A_q+A'_q$ and $<|\tilde{h}_q|^2=C_q+C'_q$. 
%As shown in Fig.~\ref{pre_factors}, the average value of $B_q$ for the data is $\simeq 0.32$. 
%As a result, the $\alpha$-helix peak in $<|\tilde{s}_q|^2>$ is 42\% larger than the peak in $A_q$, 
%and also an $\alpha$-helix peak is induced in $<|\tilde{h}_q|^2>$ as result of the peak in 
%$A_q$ even though there is no peak in $C_q$. 
%With the current value of $B_q\simeq 0.32$, the peak in $A_q$ is magnified by 42\% in $<|s_q|^2>$, 
%and a peak is induced in $<|h_q|^2>$ because of existing peak in $A_q$.

The prevalence of $\alpha$-helices in structures is reflected in the
peak at $\lambda=3.6$ in Fig.~\ref{A_q}. 
As a check, we repeated the analysis for 493 proteins in our database that
are classified as mainly $\beta$ by CATH~\cite{orengo97}.
The $\alpha$-helix peak disappears completely for this subset,
and a weaker peak corresponding to $\beta$ strands at $\lambda = 2.2$
(which was not visible in Fig.~\ref{A_q}) emerges in its places.  
This may indicate that the formation and arrangement of $\beta$ strands is
less influenced by hydrophobic forces.  The other prominent feature of
Fig.~\ref{A_q} is the increase in $A_q$ as $q\to 0$.  We believe this
reflects the fact that at a coarse level the protein is a {\em compact
  polymer}; it is well known that polymer statistics leads to
long-range correlations in the statistics of segments in the interior
of a compact structure~\cite{govorun01}.  While the precise manner in
which this could lead to correlations as in Fig.~\ref{A_q} has not
been worked out, we note that similar effects have been observed
before in studies of protein-like structures in three
dimensions~\cite{khokhlov99}, and compact lattice polymers in two
dimensions~\cite{yahyanejad03}.

The $\alpha$-helix peak, which is prominent in the hydrophobicity
power spectrum of Fig.~\ref{Hq2} is absent from Fig.~\ref{C_q}.  Thus,
the observed periodicity in sequence data is not an intrinsic feature
of the amino-acid profiles, but dictated by the required folding of
structures.  If the sequence of amino-acids were totally random, we
would expect a distribution $P(\{h_i\})=\prod_i p_a(h_i)$, where
$p_a(h_i)$ indicates the frequency of a particular base.
The corresponding distribution in Fourier space would also be
independent of $q$. The observed $\{C_q\}$ are indeed constant
(approximately $0.42\pm0.02$), at large $q$. 
This constant is different from the average indicated in Fig.~\ref{Hq2}, with the
assumption that the amino-acids are distributed randomly. This
difference is due to the interaction term in
equation~\ref{eqn:Gaussian}.

Reduced values of $C_q$ are observed as $q\to0$, corresponding to
large periodicities, as seen in Fig.~\ref{C_q}.  A similar feature is
also present in the power spectrum in Fig.~\ref{Sq2}, as noted before
by Irback {\em et al}.~\cite{irback97} who suggest that
anti-correlations can be advantageous for removing the degeneracies of
ground state for folding sequences. More recent studies also indicate
that long stretches of hydrophobic monomers, which could be a source
of long range positive correlations, are avoided~\cite{schwartz01}.
Further investigations of this issue would be helpful.

Finally, we note that the interaction terms $\{B_q\}$ in
Fig.~\ref{B_q} which correlate sequence and structure profiles (at different periodicities)
are approximately constant. 
As $\sum_q\tilde{h}_q\tilde{s_q}^*=\sum_i h_i s_i$, these terms can be
regarded as arising from the Boltzmann weight $\exp[-E/(k_B T)]$ of a
solvation energy $E=\sum_i h_i s_i$ at some temperature $T$.
Using $\overline{B_q}\approx 0.32 \pm
0.03$~kcal/mol, we can extract a corresponding temperature of $T =(2
\overline{B_q})/k_B = 323 \pm 30^{\circ} {\rm K}$.  Interestingly,
this fictitious $T$ is around room temperature, i.e. in the range of
temperatures that most proteins fold and function.  This indicates
that an important factor in correlating sequence hydrophobicity, and
structural solvent accessibility is indeed the free energy of
solvation.  This conclusion is also consistent with the analysis done
by Miller~\cite{miller87,finkelstein95}, which estimated differences in the free energies of amino-acids between the surface and the core of the
proteins by counting their relative frequencies in the different
locations. 
%Finkelstein{\em et al}~\cite{finkelstein95} provides a more
%thorough discussion on why we expect this fictitious temperature to be
%near room temperature.

%\begin{figure}[htbp]
%\begin{center}
%\psfrag{X}[l][][1.3][0] {$q$}      
%\psfrag{Y}[bc][][1.3][0]{$\chi_q$} 
%\psfrag{L}[l][][1.3][0] {$\lambda$}  
%\epsfig{file=3_17.eps, height=2.5in}
%\end{center}
%\caption{The susceptibility $\chi_q$ is negative since the 
%more hydrophobic monomers tend to  be in less solvent exposed sites. 
%The circle and square symbols correspond to real and imaginary components, respectively.
%} \label{susceptibility}
%\end{figure}

In principle, the Gaussian distribution in
Eq.~\ref{eqn:Gaussian} can be used as a tool for predicting
structures, at least as far as their surface exposure profile is concerned.
Given a specific sequence, we can calculate the hydrophobicity profile
$\{h_i\}$, and the corresponding $\{\tilde{h}_q\}$.  The conditional
probability for surface exposure profiles is then given by
\begin{equation}
\label{conditional_Gaussian}
\begin{array}{rl}
P(\{\tilde{s}_q|\tilde{h}_q\})&= \prod_q p(\tilde{s}_q|\tilde{h}_q)\propto\\
%\frac{P(s_q,h_q^\star)}{\sum_{s_q}{P(s_q,h_q^\star)}} \\
 &  \prod_{q} 
 \exp\left[-\frac{(\Re s_q-\chi_q \Re h_q)^2}{2\sigma_q^2}-\frac{(\Im s_q-\chi'_q \Im h_q)^2}{2{\sigma_q^\prime}^2}\right].\\
\end{array}
\end{equation}
Thus $\tilde{s}_q$ is Gaussian distributed with a mean value of
$\chi_q\tilde{h}_q$, and a variance $\sigma_q^2$, with the `susceptibility'
$\chi_q$, and the `noise' $\sigma_q$ easily related to
$(A_q, B_q, C_q)$.
%; $\chi_q$ is plotted in Fig.~\ref{susceptibility}.  
The corresponding distribution of
$\{s_i\}$ in real space is then obtained by Fourier transformation.

%\section*{Conclusions}
We investigated correlations between protein sequences and structures
due to hydrophobic forces, by application of Fourier transforms to
profiles of hydrophobicity and solvent accessibility.  
Each Fourier component is separately well approximated by a Gaussian distribution;
their joint distribution is described by a product of multivariate
Gaussians at different periodicities. 
This approach enables us to separate the intrinsic tendencies of the profiles 
from the interactions that couple them.  
We thus find that $\alpha$-helix periodicity is a feature of structures and
not sequences, and that at long periods the structural profiles are more
correlated than average, while the sequences are less correlated.
A quite satisfying outcome is that the correlations between the two
profiles can be explained by the Boltzmann weight of the solvation
energy at room temperatures.

Our joint distribution can be used in applications such as
predicting solvent accessibility from hydrophobicity
profiles~\cite{naderi01}, or protein interaction sites~\cite{gallet00}. 
Incorporating the impact of correlations within solvent accessibilities
is likely to improve predictions. 
The distribution can also be used in
analytical approaches to protein folding, wherever there is a need for
taking into account the complexities of structure and sequence space.

%\section*{Acknowledgments}
%{\bf Acknowledgments} 
MY and CBB are supported by Functional 
Genomics Innovation Award (CBB and Philip A. Sharp).
MK is supported by NSF through grant No. DMR-01-18213. 
MY thanks M. Povinelli for helpful discussions.

\bibliography{paper}

\begin{thebibliography}{27}
\expandafter\ifx\csname natexlab\endcsname\relax\def\natexlab#1{#1}\fi
\expandafter\ifx\csname bibnamefont\endcsname\relax
  \def\bibnamefont#1{#1}\fi
\expandafter\ifx\csname bibfnamefont\endcsname\relax
  \def\bibfnamefont#1{#1}\fi
\expandafter\ifx\csname citenamefont\endcsname\relax
  \def\citenamefont#1{#1}\fi
\expandafter\ifx\csname url\endcsname\relax
  \def\url#1{\texttt{#1}}\fi
\expandafter\ifx\csname urlprefix\endcsname\relax\def\urlprefix{URL }\fi
\providecommand{\bibinfo}[2]{#2}
\providecommand{\eprint}[2][]{\url{#2}}

\bibitem[{\citenamefont{Kauzmann}(1959)}]{kauzmann59}
\bibinfo{author}{\bibfnamefont{W.}~\bibnamefont{Kauzmann}},
  \bibinfo{journal}{Advan. Protein Chem.} \textbf{\bibinfo{volume}{16}},
  \bibinfo{pages}{1} (\bibinfo{year}{1959}).

\bibitem[{\citenamefont{Lee and Richards}(1971)}]{lee71}
\bibinfo{author}{\bibfnamefont{B.~K.} \bibnamefont{Lee}} \bibnamefont{and}
  \bibinfo{author}{\bibfnamefont{F.~M.} \bibnamefont{Richards}},
  \bibinfo{journal}{J. Mol. Biol} \textbf{\bibinfo{volume}{55}},
  \bibinfo{pages}{379} (\bibinfo{year}{1971}).

\bibitem[{\citenamefont{Eisenberg et~al.}(1984)\citenamefont{Eisenberg, Weiss,
  and Terwilliger}}]{eisenberg84}
\bibinfo{author}{\bibfnamefont{D.}~\bibnamefont{Eisenberg}},
  \bibinfo{author}{\bibfnamefont{R.~M.} \bibnamefont{Weiss}}, \bibnamefont{and}
  \bibinfo{author}{\bibfnamefont{T.~C.} \bibnamefont{Terwilliger}},
  \bibinfo{journal}{Proc. Nat. Acad. Sci. US} \textbf{\bibinfo{volume}{81}},
  \bibinfo{pages}{140} (\bibinfo{year}{1984}).

\bibitem[{\citenamefont{Moelbert et~al.}(2004)\citenamefont{Moelbert, Emberly,
  and Tang}}]{moelbert04}
\bibinfo{author}{\bibfnamefont{S.}~\bibnamefont{Moelbert}},
  \bibinfo{author}{\bibfnamefont{E.}~\bibnamefont{Emberly}}, \bibnamefont{and}
  \bibinfo{author}{\bibfnamefont{C.}~\bibnamefont{Tang}},
  \bibinfo{journal}{Protein Sci.} \textbf{\bibinfo{volume}{13}},
  \bibinfo{pages}{752} (\bibinfo{year}{2004}).

\bibitem[{\citenamefont{Irback et~al.}(1996)\citenamefont{Irback, Peterson, and
  Potthast}}]{irback96}
\bibinfo{author}{\bibfnamefont{A.}~\bibnamefont{Irback}},
  \bibinfo{author}{\bibfnamefont{C.}~\bibnamefont{Peterson}}, \bibnamefont{and}
  \bibinfo{author}{\bibfnamefont{F.}~\bibnamefont{Potthast}},
  \bibinfo{journal}{Proc. Nat. Acad. Sci. USA} \textbf{\bibinfo{volume}{93}},
  \bibinfo{pages}{9533} (\bibinfo{year}{1996}).

\bibitem[{\citenamefont{Pande et~al.}(1994)\citenamefont{Pande, Grosberg, and
  Tanaka}}]{pande94}
\bibinfo{author}{\bibfnamefont{V.~S.} \bibnamefont{Pande}},
  \bibinfo{author}{\bibfnamefont{A.~Y.} \bibnamefont{Grosberg}},
  \bibnamefont{and} \bibinfo{author}{\bibfnamefont{T.}~\bibnamefont{Tanaka}},
  \bibinfo{journal}{Proc. Nat. Acad. Sci. USA} \textbf{\bibinfo{volume}{91}},
  \bibinfo{pages}{12972} (\bibinfo{year}{1994}).

\bibitem[{\citenamefont{Strait and Dewey}(1995)}]{strait95}
\bibinfo{author}{\bibfnamefont{B.~J.} \bibnamefont{Strait}} \bibnamefont{and}
  \bibinfo{author}{\bibfnamefont{T.~G.} \bibnamefont{Dewey}},
  \bibinfo{journal}{PRE} \textbf{\bibinfo{volume}{52}}, \bibinfo{pages}{6588}
  (\bibinfo{year}{1995}).

\bibitem[{\citenamefont{Weiss and Herzel}(1998)}]{weiss98}
\bibinfo{author}{\bibfnamefont{O.}~\bibnamefont{Weiss}} \bibnamefont{and}
  \bibinfo{author}{\bibfnamefont{H.}~\bibnamefont{Herzel}},
  \bibinfo{journal}{Journal of Theoretical Biology}
  \textbf{\bibinfo{volume}{190}}, \bibinfo{pages}{341} (\bibinfo{year}{1998}).

\bibitem[{\citenamefont{Eisenhaber et~al.}(1996)\citenamefont{Eisenhaber,
  Imperiale, Argos, and Frommel}}]{eisenhaber96a}
\bibinfo{author}{\bibfnamefont{F.}~\bibnamefont{Eisenhaber}},
  \bibinfo{author}{\bibfnamefont{F.}~\bibnamefont{Imperiale}},
  \bibinfo{author}{\bibfnamefont{P.}~\bibnamefont{Argos}}, \bibnamefont{and}
  \bibinfo{author}{\bibfnamefont{C.}~\bibnamefont{Frommel}},
  \bibinfo{journal}{Protein-Struct. Funct. Genet.}
  \textbf{\bibinfo{volume}{25}}, \bibinfo{pages}{157} (\bibinfo{year}{1996}).

\bibitem[{\citenamefont{Wilder and Shakhnovich}(2000)}]{wilder00}
\bibinfo{author}{\bibfnamefont{J.}~\bibnamefont{Wilder}} \bibnamefont{and}
  \bibinfo{author}{\bibfnamefont{E.~I.} \bibnamefont{Shakhnovich}},
  \bibinfo{journal}{Phys. Rev. E} \textbf{\bibinfo{volume}{62}},
  \bibinfo{pages}{7100} (\bibinfo{year}{2000}).

\bibitem[{\citenamefont{Govorun et~al.}(2001)\citenamefont{Govorun, Ivanov,
  Khokhlov, Khalatur, Borovinsky, and Grosberg}}]{govorun01}
\bibinfo{author}{\bibfnamefont{E.~N.} \bibnamefont{Govorun}},
  \bibinfo{author}{\bibfnamefont{V.~A.} \bibnamefont{Ivanov}},
  \bibinfo{author}{\bibfnamefont{A.~R.} \bibnamefont{Khokhlov}},
  \bibinfo{author}{\bibfnamefont{P.~G.} \bibnamefont{Khalatur}},
  \bibinfo{author}{\bibfnamefont{A.~L.} \bibnamefont{Borovinsky}},
  \bibnamefont{and} \bibinfo{author}{\bibfnamefont{A.~Y.}
  \bibnamefont{Grosberg}}, \bibinfo{journal}{Phys. Rev. E}
  \textbf{\bibinfo{volume}{64}}, \bibinfo{pages}{R40903}
  (\bibinfo{year}{2001}).

\bibitem[{\citenamefont{Yahyanejad et~al.}(2003)\citenamefont{Yahyanejad,
  Kardar, and Tang}}]{yahyanejad03}
\bibinfo{author}{\bibfnamefont{M.}~\bibnamefont{Yahyanejad}},
  \bibinfo{author}{\bibfnamefont{M.}~\bibnamefont{Kardar}}, \bibnamefont{and}
  \bibinfo{author}{\bibfnamefont{C.}~\bibnamefont{Tang}}, \bibinfo{journal}{J.
  Chem. Phys.} \textbf{\bibinfo{volume}{118}}, \bibinfo{pages}{4277}
  (\bibinfo{year}{2003}).

\bibitem[{\citenamefont{Khokhlov and Khalatur}(1999)}]{khokhlov99}
\bibinfo{author}{\bibfnamefont{A.~R.} \bibnamefont{Khokhlov}} \bibnamefont{and}
  \bibinfo{author}{\bibfnamefont{P.~G.} \bibnamefont{Khalatur}},
  \bibinfo{journal}{Phys. Rev. Lett.} \textbf{\bibinfo{volume}{82}},
  \bibinfo{pages}{3456} (\bibinfo{year}{1999}).

\bibitem[{\citenamefont{Biswas et~al.}(2003)\citenamefont{Biswas, Devido, and
  Dorsey}}]{biswas03}
\bibinfo{author}{\bibfnamefont{K.~M.} \bibnamefont{Biswas}},
  \bibinfo{author}{\bibfnamefont{D.~R.} \bibnamefont{Devido}},
  \bibnamefont{and} \bibinfo{author}{\bibfnamefont{J.~G.}
  \bibnamefont{Dorsey}}, \bibinfo{journal}{J. Chromatogr. A}
  \textbf{\bibinfo{volume}{1000}}, \bibinfo{pages}{637} (\bibinfo{year}{2003}).

\bibitem[{\citenamefont{Rackovsky}(1998)}]{rackovsky98}
\bibinfo{author}{\bibfnamefont{S.}~\bibnamefont{Rackovsky}},
  \bibinfo{journal}{Proc. Nat. Acad. Sci. USA} \textbf{\bibinfo{volume}{95}},
  \bibinfo{pages}{8580} (\bibinfo{year}{1998}).

\bibitem[{\citenamefont{Irback and Sandelin}(2000)}]{irback00}
\bibinfo{author}{\bibfnamefont{A.}~\bibnamefont{Irback}} \bibnamefont{and}
  \bibinfo{author}{\bibfnamefont{E.}~\bibnamefont{Sandelin}},
  \bibinfo{journal}{Biophysical Journal} \textbf{\bibinfo{volume}{79}},
  \bibinfo{pages}{2252} (\bibinfo{year}{2000}).

\bibitem[{\citenamefont{Miller et~al.}(1987)\citenamefont{Miller, Janin, Lesk,
  and Chothia}}]{miller87}
\bibinfo{author}{\bibfnamefont{S.}~\bibnamefont{Miller}},
  \bibinfo{author}{\bibfnamefont{J.}~\bibnamefont{Janin}},
  \bibinfo{author}{\bibfnamefont{A.~M.} \bibnamefont{Lesk}}, \bibnamefont{and}
  \bibinfo{author}{\bibfnamefont{C.}~\bibnamefont{Chothia}},
  \bibinfo{journal}{J. Mol. Biol.} \textbf{\bibinfo{volume}{196}},
  \bibinfo{pages}{641} (\bibinfo{year}{1987}).

\bibitem[{\citenamefont{Finkelstein et~al.}(1995)\citenamefont{Finkelstein,
  Badretdinov, and Gutin}}]{finkelstein95}
\bibinfo{author}{\bibfnamefont{A.~V.} \bibnamefont{Finkelstein}},
  \bibinfo{author}{\bibfnamefont{A.~Y.} \bibnamefont{Badretdinov}},
  \bibnamefont{and} \bibinfo{author}{\bibfnamefont{A.~M.} \bibnamefont{Gutin}},
  \bibinfo{journal}{Protein-Struct. Funct. Genet.}
  \textbf{\bibinfo{volume}{23}}, \bibinfo{pages}{142} (\bibinfo{year}{1995}).

\bibitem[{\citenamefont{Holm and Sander}(1998)}]{holm98b}
\bibinfo{author}{\bibfnamefont{L.}~\bibnamefont{Holm}} \bibnamefont{and}
  \bibinfo{author}{\bibfnamefont{C.}~\bibnamefont{Sander}},
  \bibinfo{journal}{Nucl. Acid Res.} \textbf{\bibinfo{volume}{26}},
  \bibinfo{pages}{316} (\bibinfo{year}{1998}).

\bibitem[{\citenamefont{Orengo et~al.}(1997)\citenamefont{Orengo, Michie,
  Jones, Jones, Swindells, and Thornton}}]{orengo97}
\bibinfo{author}{\bibfnamefont{C.~A.} \bibnamefont{Orengo}},
  \bibinfo{author}{\bibfnamefont{A.~D.} \bibnamefont{Michie}},
  \bibinfo{author}{\bibfnamefont{S.}~\bibnamefont{Jones}},
  \bibinfo{author}{\bibfnamefont{D.~T.} \bibnamefont{Jones}},
  \bibinfo{author}{\bibfnamefont{M.~B.} \bibnamefont{Swindells}},
  \bibnamefont{and} \bibinfo{author}{\bibfnamefont{J.~M.}
  \bibnamefont{Thornton}}, \bibinfo{journal}{Structure}
  \textbf{\bibinfo{volume}{5}}, \bibinfo{pages}{1093} (\bibinfo{year}{1997}).

\bibitem[{\citenamefont{Fauchere and Pliska}(1983)}]{fauchere83}
\bibinfo{author}{\bibfnamefont{J.}~\bibnamefont{Fauchere}} \bibnamefont{and}
  \bibinfo{author}{\bibfnamefont{V.}~\bibnamefont{Pliska}},
  \bibinfo{journal}{Eur. J. Med. Chem.} \textbf{\bibinfo{volume}{18}},
  \bibinfo{pages}{369} (\bibinfo{year}{1983}).

\bibitem[{\citenamefont{Hubbard et~al.}(1991)\citenamefont{Hubbard, Campbell,
  and Thornton}}]{hubbard91}
\bibinfo{author}{\bibfnamefont{S.~J.} \bibnamefont{Hubbard}},
  \bibinfo{author}{\bibfnamefont{S.~F.} \bibnamefont{Campbell}},
  \bibnamefont{and} \bibinfo{author}{\bibfnamefont{J.~M.}
  \bibnamefont{Thornton}}, \bibinfo{journal}{J. Mol. Biol.}
  \textbf{\bibinfo{volume}{220}}, \bibinfo{pages}{507} (\bibinfo{year}{1991}).

\bibitem[{\citenamefont{Weikl and Dill}(2003)}]{weikl03}
\bibinfo{author}{\bibfnamefont{T.~R.} \bibnamefont{Weikl}} \bibnamefont{and}
  \bibinfo{author}{\bibfnamefont{K.~A.} \bibnamefont{Dill}},
  \bibinfo{journal}{J. Mol. Biol.} \textbf{\bibinfo{volume}{332}},
  \bibinfo{pages}{953} (\bibinfo{year}{2003}).

\bibitem[{\citenamefont{Irback et~al.}(1997)\citenamefont{Irback, Peterson, and
  Potthast}}]{irback97}
\bibinfo{author}{\bibfnamefont{A.}~\bibnamefont{Irback}},
  \bibinfo{author}{\bibfnamefont{C.}~\bibnamefont{Peterson}}, \bibnamefont{and}
  \bibinfo{author}{\bibfnamefont{F.}~\bibnamefont{Potthast}},
  \bibinfo{journal}{Phys. Rev. E.} \textbf{\bibinfo{volume}{55}},
  \bibinfo{pages}{860} (\bibinfo{year}{1997}).

\bibitem[{\citenamefont{Schwartz et~al.}(2001)\citenamefont{Schwartz, Istrail,
  and King}}]{schwartz01}
\bibinfo{author}{\bibfnamefont{R.}~\bibnamefont{Schwartz}},
  \bibinfo{author}{\bibfnamefont{S.}~\bibnamefont{Istrail}}, \bibnamefont{and}
  \bibinfo{author}{\bibfnamefont{J.}~\bibnamefont{King}},
  \bibinfo{journal}{Protein Sci.} \textbf{\bibinfo{volume}{10}},
  \bibinfo{pages}{1023} (\bibinfo{year}{2001}).

\bibitem[{\citenamefont{Naderi-manesh et~al.}(2001)\citenamefont{Naderi-manesh,
  Sadeghi, Arab, and Movahedi}}]{naderi01}
\bibinfo{author}{\bibfnamefont{H.}~\bibnamefont{Naderi-manesh}},
  \bibinfo{author}{\bibfnamefont{M.}~\bibnamefont{Sadeghi}},
  \bibinfo{author}{\bibfnamefont{S.}~\bibnamefont{Arab}}, \bibnamefont{and}
  \bibinfo{author}{\bibfnamefont{A.~A.} \bibnamefont{Movahedi}},
  \bibinfo{journal}{Protein-Struct. Funct. Genet.}
  \textbf{\bibinfo{volume}{42}}, \bibinfo{pages}{452} (\bibinfo{year}{2001}).

\bibitem[{\citenamefont{Gallet et~al.}(2000)\citenamefont{Gallet, Charloteaux,
  Thomas, and Brasseur}}]{gallet00}
\bibinfo{author}{\bibfnamefont{X.}~\bibnamefont{Gallet}},
  \bibinfo{author}{\bibfnamefont{B.}~\bibnamefont{Charloteaux}},
  \bibinfo{author}{\bibfnamefont{A.}~\bibnamefont{Thomas}}, \bibnamefont{and}
  \bibinfo{author}{\bibfnamefont{R.}~\bibnamefont{Brasseur}},
  \bibinfo{journal}{J. Mol. Biol.} \textbf{\bibinfo{volume}{302}},
  \bibinfo{pages}{917} (\bibinfo{year}{2000}).

\end{thebibliography}

\end{document}